\pdfoutput=1
\documentclass[sigconf]{acmart}
\usepackage[english]{babel}
\renewcommand\footnotetextcopyrightpermission[1]{}
\setcopyright{none}
\makeatletter
\def\@copyrightpermission{}
\makeatother
\acmDOI{}

\acmISBN{}

\acmPrice{}
\usepackage{printlen}
\usepackage{hyperref}
\usepackage[small,compact]{titlesec}
\usepackage{mathtools}
\usepackage{multirow}
\usepackage{wrapfig}
\usepackage{url}
\usepackage{xcolor}
\usepackage{enumitem}
\usepackage{capt-of}
\usepackage{booktabs}
\usepackage{subcaption}
\usepackage{calrsfs}
\usepackage{listings}
\usepackage{amsfonts}

\usepackage[linesnumbered,noend]{algorithm2e}
\usepackage{etoolbox}
\providetoggle{beginFlag}
\settoggle{beginFlag}{true}

\SetAlgoSkip{SkipBeforeAndAfter}

\usepackage{xspace}

\usepackage[noabbrev]{cleveref}

\newcommand{\eg}{{\it e.g.}, }
\newcommand{\ie}{{\it i.e.}, }

\newcommand{\system}{\textsc{Harmonia}\xspace}

\usepackage{pifont} 
\newcommand{\cmark}{\ding{51}}%
\newcommand{\xmark}{\ding{55}}%

\usepackage{graphicx}

\definecolor{brewerpurple}{HTML}{AF4EA3}
\definecolor{brewerblue}{HTML}{377EB8}
\definecolor{NavyBlue}{HTML}{006EB8}
\definecolor{BrickRed}{HTML}{B6321C}
\definecolor{ForestGreen}{HTML}{009B55}
\lstdefinestyle{customc2}{
    emph={update\_metrics, prune\_completed\_jobs, exec\_jobs, accept, schedule, place, AcceptAll, Fifo, Consolidated, pop\_wait\_queue, update\_cluster},
    escapeinside={(*@}{@*)},
    emphstyle=\bfseries\color{NavyBlue},
    commentstyle=\color{ForestGreen}\itshape\ttfamily,
       morekeywords={yield},
    stringstyle=\color{red}\ttfamily,
    emph={[2]admission\_policy, scheduling\_policy, placement\_policy},emphstyle={[2]\bfseries\color{BrickRed}},
    language=Python,
    keywordstyle=\bfseries\color{green!40!black}
}

\begin{document}
\title{\system: End-to-End RAG Serving Optimization}

\author{Saurabh Agarwal}
\authornote{Both authors contributed equally to this research.}
\affiliation{%
  \institution{UT Austin}
}

\author{Bodun Hu}
\authornotemark[1]
\affiliation{%
  \institution{UT Austin}
}

\author{Luis Pabon}
\affiliation{%
  \institution{UT Austin}
}

\author{Myungjin Lee}
\affiliation{%
  \institution{Cisco Research}
}

\author{Jayanth Srinivasa}
\affiliation{%
  \institution{Cisco Systems}
}

\author{Aditya Akella}
\affiliation{%
  \institution{UT Austin}
}

\begin{abstract}

Retrieval-Augmented Generation (RAG) improves LLM reliability by integrating external knowledge, but serving RAG efficiently is challenging because requests traverse heterogeneous pipelines spanning LLM inference, databases, and CPU-side processing. We present \system{}, an end-to-end RAG serving framework that addresses these bottlenecks via (i) a flexible pipeline specification interface for composing custom workflows, (ii) heterogeneity-aware deployment that provisions and configures components as a distributed inference system, and (iii) a closed-loop runtime controller that monitors load and execution progress and reduces SLO violations through request prioritization and auto-scaling. Across four RAG applications, \system{} outperforms commercial alternatives, improving throughput by more than $2.04\times$ while reducing SLO violations by up to $78.4\%$.
\end{abstract}

\maketitle
\thispagestyle{plain}
\pagestyle{plain}



\section{Introduction}
Compound AI applications are emerging as the next frontier of distributed, networked systems. Rather than issuing a single model call, they execute multi-stage workflows that combine retrieval over external knowledge, query transformation, and LLM inference, often with iterative refinement and tool use. Retrieval-Augmented Generation (RAG) is a widely deployed instance of this class: each query induces a request graph spanning vector search, data stores, CPU stages, and GPU-backed decoding, with intermediate state and token streams flowing across machines and racks. This paper seeks a general framework for high-throughput, low-latency execution of such RAG workflows at scale, despite rapidly evolving components and dynamic workflow structure.

Just as RPC-based microservices must identify end-to-end critical paths and coordinate control across services to achieve good performance~\cite{crisp, 40801, 36356, 10.1145/3570522}, RAG pipelines now demand workflow-wide management. Their performance is governed by cross-stage sequences of retrieval, computation, data movement, and queuing effects. 

Delivering predictable performance for RAG is difficult for reasons intrinsic to these workflows, with Section~\ref{sec:background} providing quantitative evidence. First, \emph{the component stack evolves rapidly}: retrieval, augmentation, and generation are all in flux, and modern pipelines increasingly incorporate safety gates, efficiency layers, and adaptive control flow. Rigid framework abstractions become costly under this pace of change because they turn routine pipeline edits into framework-level engineering. For example, Section~\ref{sec:background} shows that expressing common RAG logic in widely used stacks often requires framework-specific operators instead of standard Python control flow, and integrating new components can require modifications to the serving framework itself.

Second, \emph{resource usage is heterogeneous, and bottlenecks move with workflow structure}. RAG mixes CPU-, memory-, and I/O-bound stages with GPU-resident decoding, and the fraction of end-to-end time spent in retrieval can vary widely across representative workflows (Figure~\ref{fig:performance_hetero}). As a result, a deployment tuned for one workflow can be misprovisioned for another, even when using the same underlying components. Yet many popular stacks deploy RAG as monolithic Python processes, scaling the entire workflow as a single unit. This forces all stages to share one resource pool and makes coarse-grained replication the primary scaling knob, even when only one stage is the true bottleneck.

Third, \emph{execution structure and cost are request-dependent}, undermining static provisioning and per-component heuristics. Control flow can branch, recurse, or exit early based on intermediate results, and per-stage work varies with query complexity, retrieval depth, and generation length. This stochasticity also interacts with cross-stage communication mechanisms in subtle ways. For instance, streaming is commonly used to overlap upstream compute with downstream prefill, but Section~\ref{sec:background} shows that its efficacy is load-dependent: it can improve performance at low load, yet reduce throughput and degrade performance at high load due to pipeline stalls caused by unmanaged resource contention (Figure~\ref{fig:streaming_impact}). These effects remain invisible to narrow runtime interfaces that only expose per-component metrics and lack workflow-wide telemetry.

To address these challenges, we advocate that the RAG workflow be exposed as a high-level, executable request graph to a control plane that sets graph-level knobs, while separating slow-timescale planning from fast-timescale adaptation. Specifically, we enable end-to-end coordination among (i) a programming interface that preserves developer flexibility while still yielding a machine-readable control flow representation, (ii) an optimization layer that auto-provisions heterogeneous resources at the component granularity reflected in that representation, and (iii) a runtime layer that uses graph-level telemetry to adapt workflow-wide routing, etc., under variance. Addressing any one of these in isolation is insufficient. Without a representation that captures evolving components and dynamic control flow, the system cannot reason about dependencies or critical paths. Without component-aware provisioning, it over-allocates to non-bottleneck stages, forcing repeated manual retuning as workflows evolve. Without runtime visibility and control, deployment-time choices deteriorate under load shifts and query-dependent behavior, leading to queueing cascades and pipeline stalls (as illustrated in Figure~\ref{fig:streaming_impact}).

This paper presents \system{}, a RAG serving engine built around the above ideas. In \system{}, developers write rich pipelines in idiomatic Python as if they are targeting a single logical node deployment. \system{} captures the resulting computational graph from the running program. This avoids forcing developers into a brittle DSL while preserving the inter-stage dependencies that can aid both deployment planning and runtime control. 

\system{}'s deployment layer leverages the captured dependency structure to model the workflow as a directed graph with profile-driven conditional paths, accounting for heterogeneous, multi-dimensional, and data-dependent resource needs as well as cross-stage routing. It provisions resources in the underlying distributed environment by jointly solving for placement and per-stage capacity while auto-generating the communication substrate. This accommodates high-level objectives while shielding developers from low‑level deployment details.

\system{} complements this deployment-time plan with a closed-loop runtime layer that enforces objectives under run-time variance. A controller tracks execution progress and queueing at graph nodes, uses node-level resource pressure signals to trigger targeted autoscaling, and performs load-aware routing and prioritization when SLO violation risk increases. To keep this lightweight, \system{} decouples control from data movement: the controller makes scheduling decisions while intermediate results flow directly between producer and consumer stages, and data is routed back through the controller only when required by the program's control flow. Finally, \system{} provides a managed streaming primitive that standardizes buffering, chunking, and readiness signaling across stages, enabling low-latency pipelining and performance management without embedding ad-hoc stream management into application code.
\system is implemented in \textasciitilde 6K lines of Python, using gRPC as the underlying communication protocol for low-latency, high-performance messaging between the control plane and distributed workers.
To support the research community and facilitate further innovation in agentic orchestration, we will completely open-source \system{}.
We evaluate \system on four different types of popular RAG architectures and compare it against two widely used commercial frameworks, LangChain~\cite{Chase_LangChain_2022} and Haystack~\cite{Pietsch_Haystack_the_end-to-end_2019}. \system{} provides up to $2.04\times$ higher throughput (avg $1.6\times$) and reduce SLO violations by up to $78.4\%$.


\section{Background and Motivation}
\label{sec:background}

\begin{figure*}[t]
    \centering
    \includegraphics[width=0.8\textwidth]{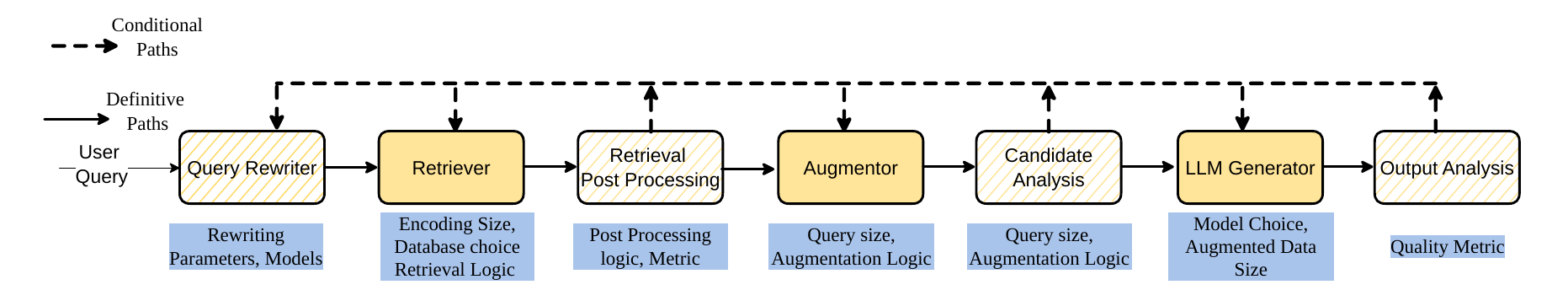}
    \vspace{-10pt}
    \caption{\textbf{Components of a RAG pipeline:} 
    The schematic illustrates the execution flow of a RAG pipeline. Solid boxes represent mandatory components, while hatched boxes indicate optional ones. The broken lines indicate potentially conditional paths. The labels beneath the boxes highlight a subset of parameters that affect scalability.}
    \label{fig:rag_pipeline}
    \vspace{-15pt}
\end{figure*}

RAG is a particularly important instance of an emerging class of \emph{compound AI applications}: user requests execute as multi-stage workflows that combine retrieval, transformation, and model inference, often with data-dependent control flow. These applications expose a central tension for system design: developers need the programming flexibility to rapidly evolve pipeline structures and components, while operators need enough workflow-level visibility and control to manage throughput and tail latency under highly variable per-request costs. We use RAG as a concrete, widely deployed representative of this class to ground the abstractions and mechanisms developed in this paper.

\subsection{Retrieval-Augmented Generation}

Retrieval-Augmented Generation (RAG)~\cite{lewis2020pre} enhances the capabilities of a Large Language Model (LLM) by supplementing it with an external, indexed knowledge base. In a typical RAG setup, a user query triggers the retrieval of relevant documents or data from this external source. The retrieved content is incorporated into a prompt and passed to the LLM (or another generative model, such as a multi-modal model) to generate a response. This augmentation improves accuracy and mitigates hallucinations.
Unlike standard LLM serving~\cite{kwon2023efficient, agrawal2023sarathi, mei2024helix}, RAG systems involve multiple components. 
Figure~\ref{fig:rag_pipeline} illustrates the diverse components a query might traverse in a RAG workflow, spanning retrieval, optional reranking and filtering, prompt construction, and GPU-backed decoding. 

\subsection{RAG Serving Challenges}
\label{sec:challenges}
Efficient RAG serving faces three stack-wide challenges in the \emph{programming model}, \emph{deployment}, and \emph{runtime control}.

\noindent{\em \textbf{\#1 Rigid Programming Model Impedes RAG development.} }
Originally, RAG architectures were static pipelines~\cite{lewis2020pre}. However, recent designs have evolved into heterogeneous, multi-stage workflows~\cite{wang2024richrag,martis2024luminirag,singh2024chunkrag,xiao2024tkg,guo2020accelerating}. Crucially, each stage is undergoing substantial and rapid evolution: 
Retrieval has expanded beyond plain-text corpora~\cite{lewis2020pre,jin2024flashrag,wang2024richrag} to include diverse structures such as vector embeddings~\cite{qian2024memorag,hipporag}, graphs~\cite{edge2024local,han2024retrieval}, trees~\cite{li2025cft,fatehkia2024t,liu2025hoprag}, and bitmaps~\cite{chen2024hiqa}. In augmentation, simple concatenation~\cite{borgeaud2022improving} has been replaced with learned ranking and filtering using classifiers or LLMs~\cite{shao2024scaling}.
In generation, state-space~\cite{gu2023mamba,fu2022hungry} and diffusion models~\cite{chandrasegaran2025exploring} now complement transformer-based ones. Furthermore, these components are increasingly being composed into rich, conditional workflows incorporating query preprocessing~\cite{martis2024luminirag,lewis2020pre}, safety gating~\cite{liang2025saferag,mukherjee2025sosecure,wu2024medical}, efficiency layers~\cite{singh2024chunkrag,xiao2024tkg,guo2020accelerating}, and adaptive control flow~\cite{yan2024corrective,asai2024self,jeong2024adaptive}.

\begin{figure}
\centering
\resizebox{1.0\linewidth}{!}{
\begin{lstlisting}[style=customc2]
from langchain_chroma import Chroma
from langchain_core.output_parsers import StrOutputParser
from langchain_core.prompts import PromptTemplate
from langchain_core.runnables import RunnablePassthrough
from langchain_openai import OpenAIEmbeddings, ChatOpenAI
from langchain_text_splitters import CharacterTextSplitter
from langchain_community.llms import VLLM

# Get Embeddings Model
embeddings = OpenAIEmbeddings(model="text-embedding-3-large")
# Initialize ChromaDB as Vector Store
vector_store = Chroma(
    collection_name="test_collection",
    embedding_function=embeddings
)
def format_docs(docs):
    return "\n\n".join(doc.page_content for doc in docs)
retriever = vector_store.as_retriever()
llm = VLLM(
    model="mosaicml/mpt-7b",
    trust_remote_code=True,  # mandatory for hf models
    max_new_tokens=128,
    top_k=10,
)
prompt_template = """My very custom prompt
context: {context}
question: {query}
answer: """
# Create Prompt Instance from template
custom_rag_prompt = PromptTemplate.from_template(prompt_template)
# Chain elements to create a RAG Graph
rag_chain = (
    {"context": retriever | format_docs, "query": RunnablePassthrough()}
    | custom_rag_prompt
    | llm
    | StrOutputParser()
)
rag_chain.invoke("Where is Hawaii?")
\end{lstlisting}}
\vspace{-10pt}
\caption{\small{\textbf{Building a RAG in LangChain:} The code above constructs a RAG pipeline using LangChain. In order to implement new abstractions, developers must modify LangChain itself. Moreover, LangChain does not provide any deployment or runtime support.}}
\label{fig:langchain_code}
\vspace{-15pt}
\end{figure}

However, integrating these advances into workflows is challenging in existing RAG frameworks (LangChain~\cite{Chase_LangChain_2022} and Haystack~\cite{Pietsch_Haystack_the_end-to-end_2019}) due to their rigid declarative programming model. To make this concrete, Figure~\ref{fig:langchain_code} shows an example of programming RAG using Langchain. To construct a RAG workflow, developers need to use LangChain-specific graph operators (e.g., RunnablePassthrough) rather than standard Python control flow. Also, to use these Langchain-specific operators effectively, all modules in the pipeline need to be implemented in LangChain internal library. Thus, if a developer needs a new type of retriever, they will need to {\em modify the internals of LangChain or wait for LangChain maintainers to add support}.

This undermines rapid experimentation and exploration of alternative RAG workflows\footnote{Developers lament that Langchain abstractions hurt productivity~\cite{langchain-critical}}. 
Ideally, to ensure modular extensibility, we need a framework with a unified programming interface that accommodates emerging components without requiring internal framework modifications.

\begin{figure*}[t]
    \centering
    \begin{minipage}[t]{0.32\linewidth}
    \vspace{-5pt}
    \includegraphics[width=\linewidth]{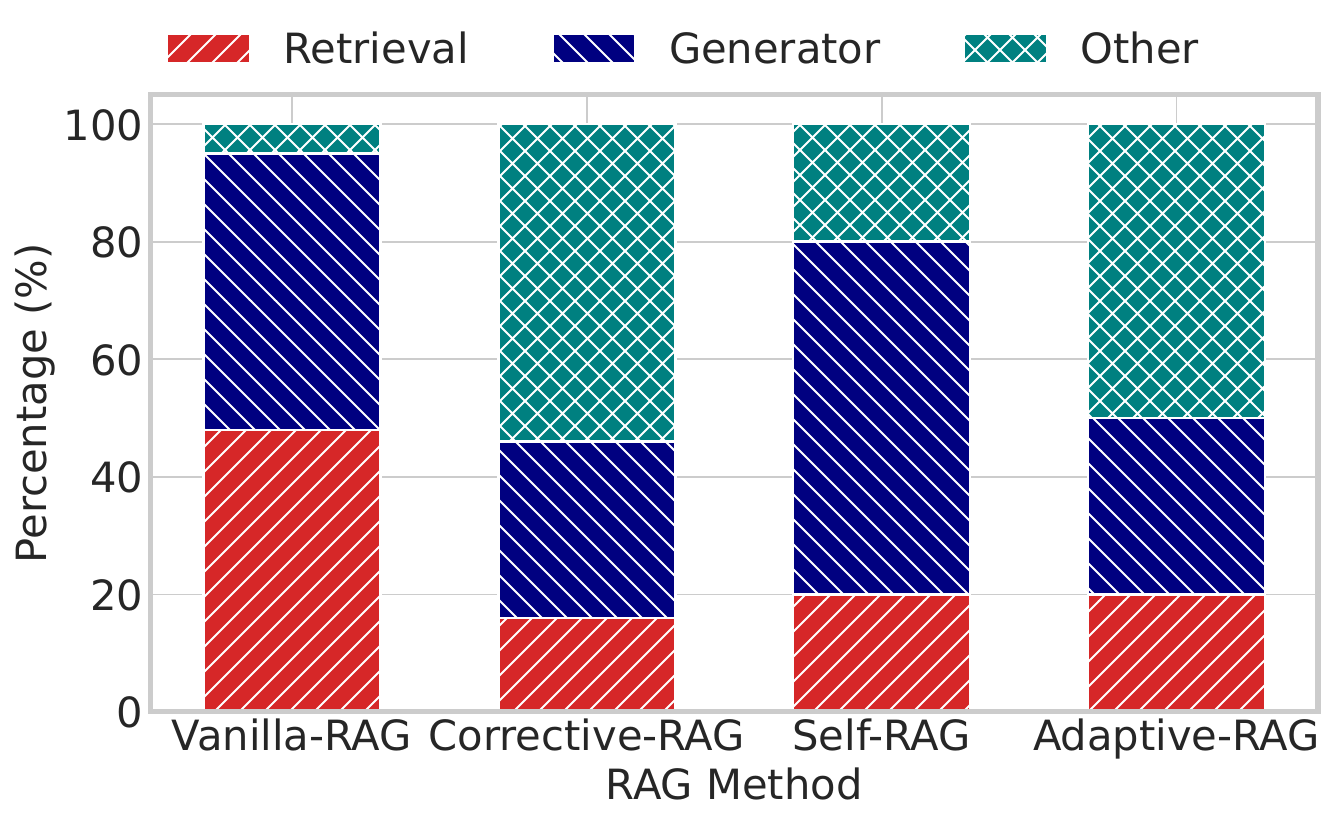}
    \caption{\small{{Performance Heterogeneity:} Average time spent in each component across four RAG workflows under identical load and datasets. The results show that different RAG pipelines exhibit distinct performance bottlenecks.
 }}
    \label{fig:performance_hetero}
    \end{minipage}\quad
    \begin{minipage}[t]{0.32\linewidth}
    \vspace{-5pt}
    \includegraphics[width=\linewidth]{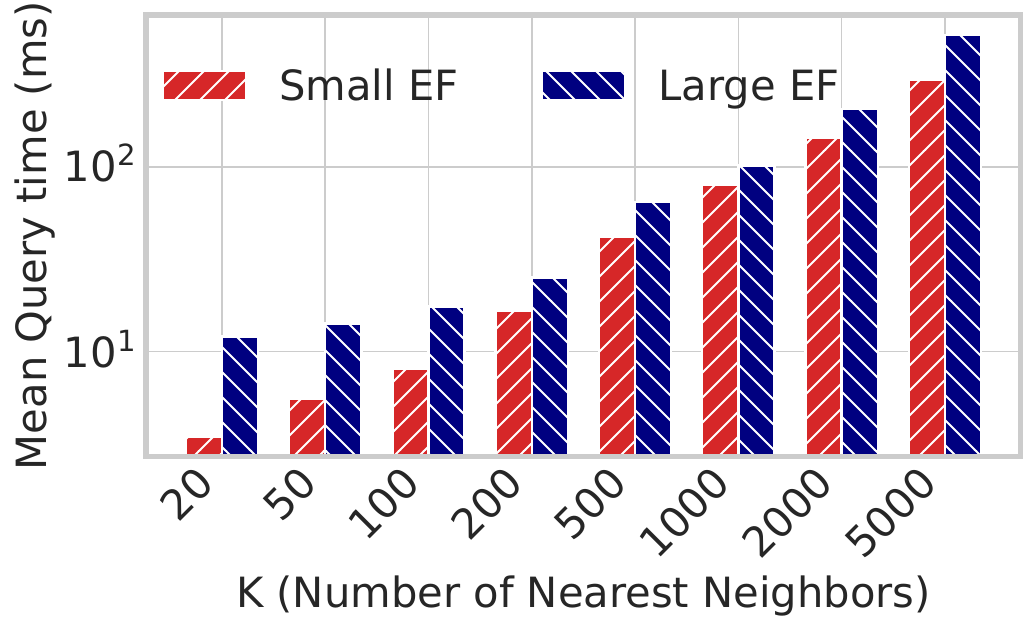}
    \caption{\small{Shifting Optimal Resource allocation: In ChromaDB, retrieval latency and accuracy are tuned via the \texttt{search\_ef} parameter. We observe, for small K, low \texttt{search\_ef} values can be up to $20\times$ faster. }}
    \label{fig:retrieval_time}
    \end{minipage}\quad
    \begin{minipage}[t]{0.32\linewidth}
    \vspace{-5pt}
    \includegraphics[width=\linewidth]{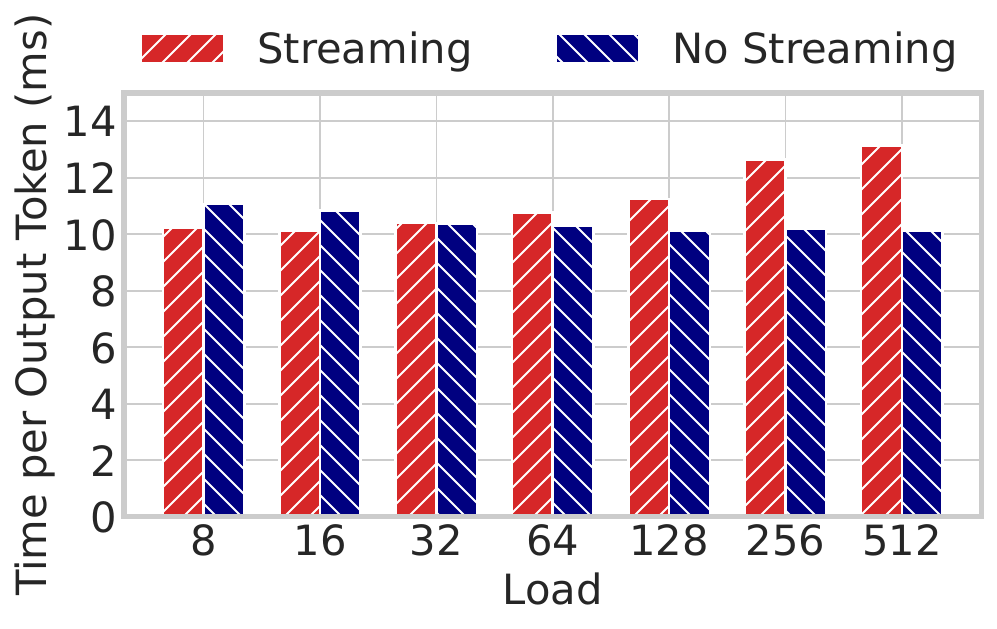}
    \caption{\small{{Impact of Streaming:} We observe that streaming improves performance at lower loads by over ~11\% while at high loads it leads to a degradation of ~24\%. }}
    \label{fig:streaming_impact}
    \end{minipage}\quad
    \vspace{-20pt}
\end{figure*}

\noindent{\em \textbf{\#2 Monolithic View Leads to poor resource utilization.}} 
RAG workflows exhibit extreme resource and performance heterogeneity. Retrieval stages are primarily CPU-, memory-, and I/O-bound, while generation stages are compute-intensive and almost exclusively GPU-resident. Our analysis (Figure~\ref{fig:performance_hetero}) reveals that the performance bottleneck is a moving target: retrieval can account for anywhere from 18\% to 62\% of end-to-end latency depending on the workflow topology. Furthermore, performance is highly sensitive to component-specific configurations; as shown in Figure~\ref{fig:retrieval_time}, tuning retrieval parameters (e.g., \texttt{search\_ef} in ChromaDB) can drastically shift the runtime performance profile of a single component.

Thus, a RAG serving system must be bottleneck-aware, providing the ability to independently scale components based on their specific resource demands. For instance, a Graph RAG pipeline might require a 3:1 ratio of retrievers (62\% of runtime) to generators (23\% of runtime) to maintain maximal throughput. However, existing frameworks like Haystack~\cite{Pietsch_Haystack_the_end-to-end_2019} and LangChain~\cite{langchain-critical} treat RAG workflows as monolithic Python processes. This architectural choice forces all components into a shared resource bucket, leaving "coarse-grained replication" as the only scaling mechanism.

Ideally, RAG deployments require an engine capable of fine-grained resource orchestration, automatically synthesizing optimal allocations to maximize both hardware efficiency and end-to-end performance.

\noindent{\em \textbf{\#3 No Visibility and Control Impede Meeting Run-time Goals Under Stochasticity.}}
Beyond static overheads—such as index scale and component selection—RAG workloads exhibit significant runtime stochasticity. Resource demands and execution latencies are highly sensitive to input-dependent variables, including query complexity, iterative retrieval cycles, and generation length. Modern pipelines further exacerbate this through data-dependent control flow; for instance, Corrective RAG~\cite{yan2024corrective} may trigger recursive retrieval based on verification results, while adaptive variants dynamically scale retrieval operations based on query difficulty~\cite{jeong2024adaptive, su2025fast}.

This stochasticity also affects inter-component communication. Although streaming is often used to overlap upstream computation with downstream pre-fill, its effectiveness is highly load-dependent. As shown in Figure~\ref{fig:streaming_impact}, streaming at high load reduces throughput by 36\%, indicating that unmanaged streaming can preempt active decoding and introduce pipeline stalls.

These characteristics necessitate a runtime management layer capable of continuous, closed-loop orchestration. To maximize throughput and ensure SLO compliance, such a layer must maintain global visibility into request state and resource telemetry, dynamically adjusting resource allocations, request priorities, and communication granularity in real-time. However, existing frameworks lack the necessary internal hooks and visibility to offer this adaptation. 

An ideal RAG serving system must decouple the functional workflow from a policy-driven execution engine, enabling sophisticated management strategies that adapt to the inherent volatility of RAG workloads.

\section{\system}

\begin{figure}[t]
    \begin{center}
        \begin{subfigure}[b]{0.45\linewidth}
        \includegraphics[width=\linewidth]{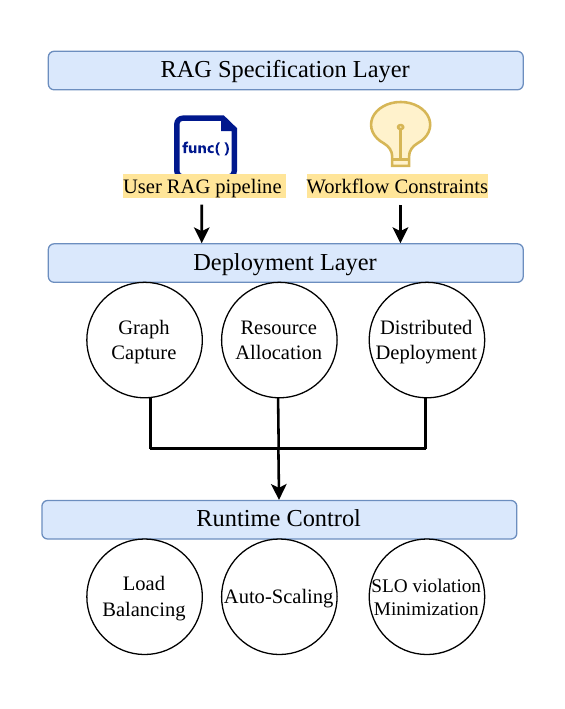}
        \caption{\small{\system workflow}}
        \end{subfigure}
        \vspace{-5pt}
    \begin{subfigure}[b]{0.35\linewidth}
     \includegraphics[width=\linewidth]{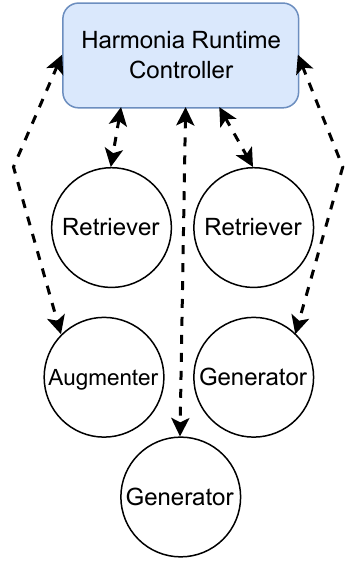}
     \caption{\small{Runtime}}
    \end{subfigure}
    \vspace{-5pt}
    \end{center}
    \caption{\small{\textbf{\system overview:} \system takes a RAG specification, transforms it into a computational graph, and distributes its nodes across multiple workers. A runtime controller coordinates requests between workers. }}
    \vspace{-15pt}
    \label{fig:patchwork_overview}
\end{figure}


\label{sec:design_goals}

\noindent
\textbf{Goals and Overview:}
\system{} is an end-to-end serving system for generic RAG pipelines, organized as a three-layer stack (Figure~\ref{fig:patchwork_overview}) that targets three requirements: (i) \emph{flexibility in workflows}: developers can add or swap components with minimal code changes; (ii) \emph{heterogeneity awareness}: serving automatically allocates and deploys CPU/GPU resources across stages; and (iii) \emph{dynamic performance control}: to manage scaling, cross-stage communication, and SLO compliance under query-dependent variance. To this end, the \emph{workflow specification layer} captures an executable pipeline graph from idiomatic Python and exposes modular component interfaces; the \emph{deployment layer} computes placement, replication, and pipeline-wide configurations to maximize throughput; and the \emph{runtime layer} monitors progress and adapts routing, scaling, and communication to address changing bottlenecks and limit tail latency.

\subsection{The RAG Specification Layer}
\label{sec:rag_spec}

The specification layer provides an intent-driven interface for composing RAG pipelines while exposing the framework hooks needed to instantiate components and manage cross-component communication. Its design emphasizes:
(i) \emph{Library-agnostic component integration:} developers can implement components using any external library, subject to minimal framework constraints.
(ii) \emph{High-level guidance without low-level control:} developers may supply coarse hints (\eg number of component instances), while \system{} determines resource allocation, scheduling, and communication interfaces to distribute execution across nodes.

Prior systems~\cite{aiken2022programming, paszke2017automatic, zaharia2010spark, meng2016mllib} for specifying compute graphs often rely on domain-specific languages (DSLs), which add friction by forcing developers to adopt a new programming model. Moreover, as discussed in \S~\ref{sec:challenges}, the declarative interfaces used in many commercial RAG systems require the full execution graph to be specified upfront, an assumption that breaks for RAG workflows with loops, conditionals, and other data-dependent control flow.

To build \system{}'s interface, we start from Ray's programming model~\cite{moritz2018ray} but make principled customizations for RAGs. Like Ray, developers annotate RAG components with a decorator. Beyond this, \system's programming model diverges from Ray. Unlike Ray, where developers need to explicitly start long running stateful actors ("detached actors" in Ray's terms, \ie the Ray scheduler cannot kill those actors or start more actors), \system{} treats each decorated class as a fully managed distributed actor -- each actor is long-running, with its launch and resource allocation explicitly managed by \system. This is a better fit for RAGs, where components are stateful and incur significant initialization and cold-start costs. \system{} thus provides automatic resource allocation and elastic scaling, allowing components to retain state and adapt to query load across single-node and multi-node deployments without any changes to user code. Figure~\ref{fig:code_listing} shows a reference implementation of C-RAG~\cite{yan2024corrective}.

The programming model incorporates three additional components that decouple intent from execution details. 

\noindent
{\bf Streaming Object}
Current distributed runtimes (e.g., Ray \cite{moritz2018ray}) and RAG-specific frameworks (e.g., LangChain~\cite{Chase_LangChain_2022}, Haystack\cite{Pietsch_Haystack_the_end-to-end_2019}, Ayo~\cite{tan2025towards}) either lack cross-component streaming support or impose a significant implementation burden on the user to manage stream lifecycles (\S\ref{sec:challenges}).
 \system introduces a managed Streaming Object abstraction, decoupling the developer's logic from the complexities of low-latency communication and synchronization. 
 As illustrated in Figure~\ref{fig:code_listing} (Line 11), a developer initializes the object and issues writes at any frequency.
 The underlying runtime intercepts these calls, 
 and controls the communication granularity  to meet end-to-end workflow goals (\S\ref{sec:runtiume_control}).
 

\noindent
{\bf Serving-Ready Classes.}
\system provides a library of base classes for common RAG primitives, including Retriever, Generator, and Augmenter. 
As illustrated in Figure~\ref{fig:code_listing}, developers implement a custom inference function by inheriting from the base class. These {\em serving-ready} abstractions automatically handle essential systems-level "book-keeping"—such as request-ID tracking, state management, and metadata propagation—relieving the developer from the complexities of implementing request lifecycle management for each new type of component.
Developers can easily extend these classes to implement new components (Line 7 and 17 in Figure~\ref{fig:code_listing}).


\noindent
{\bf Specifying workflow constraints.} 
RAG workflows often impose constraints that developers may wish to express explicitly. \system allows developers to specify such constraints directly when defining workflows, enabling runtime enforcement without additional control logic. Currently, \system supports three classes of constraints.
First, developers can mark components as \emph{stateful}. In recursive workflows, all invocations of a stateful component must be routed to the same instance to preserve per-request state. By declaring a component as stateful (Line 16, Fig.~\ref{fig:code_listing}), \system automatically ensures consistent routing.
Second, developers can specify \emph{resource requirements} (Line 5, Fig~\ref{fig:code_listing}) for each component, such as CPU, GPU, and memory demands. These constraints are fed into our resource allocation algorithm(\S~\ref{sec:deployment}). The resource allocation accounts for these constraints when performing scaling, \ie it only instantiates another copy of a component, if all the resource constraints are met. 
Finally, developers can configure a minimum number of \emph{base instances} (Line 16, Fig~\ref{fig:code_listing}) for a component. This prevents under-provisioning of infrequently invoked components with high startup costs, ensuring instances are available when requests arrive.
Together, these constraints let developers guide system behavior declaratively, without writing explicit scheduling or resource-management code.


\begin{figure}
\centering
\resizebox{0.95\linewidth}{!}{
\begin{lstlisting}[style=customc2,escapechar=|]
import harmonia
from harmonia import Augmenter, Retriever, VLLM, WebSearch, Rewriter

########### 1. Initializing harmonia ##########
@harmonia.make(base_instances=1, resources={"GPU":1, "CPU":4})
# creating a streaming Retriever
class StreamingRetriever(Retriever):
    def __init__(self):
        super().__init__()
    def retrieve(self, input):
        return_stream = harmonia.stream 
        # custom retriever writes to return_stream 
        custom_retriever(input, output=return_stream)
        return return_stream
# creating Grader as subclass of VLLM
@harmonia.make(base_instances=2, stateful=True)
class Grader(VLLM):
    def __init__(self, sys_prompt):
        super().__init__()
        self.sys_prompt = sys_prompt
    def grade(self, data):
        return self.generate(self.sys_prompt + data)

# initialize components
retriever = StreamingRetriever()
grader = Grader(sys_prompt='Does retrieved docs have relevant info?')
web = harmonia.make(WebSearch(output_format=list))
llm = harmonia.make(LLM(temp=0.7), stateful=True)

########### 2. Building computational graph ###########
# mark graph capture beginning
def main(query):
    with harmonia.capture():
        retrieved_data = retriever.retrieve(query)
        has_relevant_doc = grader.generate(retrieved_data)
        if not has_relevant_doc:
            better_query = rewriter.custom_inference(query)
            doc = web.search(better_query)
            return llm.generate(better_query, doc)
        else:
            return llm.generate(query, retrieved_data)

########### 5. Automatic RAG deployment ###########
if __name__ == "__main__":
    query = "What is the latest version of Linux Kernel"
    inference_endpoint = harmonia.deploy(main())
    inference_endpoint(query)
\end{lstlisting}}
\vspace{-10pt}
\caption{\small{\textbf{\system flow:} The code above code shows the implementation of CRAG in \system. The example highlights that \system doesn't inhibit developers' expressivity. The developer uses \system provided utilities to convert their Python workflow to efficient deployment.}}
\label{fig:code_listing}
\vspace{-15pt}
\end{figure}

\subsection{Deployment Layer}
\label{sec:deployment}
\label{sec:resource_allocation}
Deploying RAG workflows onto underlying infrastructure to optimize performance presents {\em a fundamental tension between imperative and declarative execution models}. Imperative frameworks (\eg Ray~\cite{moritz2018ray}, CIEL~\cite{ciel}) treat workloads as opaque, unstructured tasks, which ensures programmer flexibility but inhibits proactive resource planning and cross-component optimization. Conversely, declarative systems (\eg Spark~\cite{zaharia2010spark}, Dryad~\cite{dryad}, Naiad~\cite{naiad}) require a fully specified, static data-flow graph to perform optimizations; this assumption is incompatible with the data-dependent conditionals and iterative loops in RAG pipelines.

One of our core insights is that, although \system's programming model is imperative, we can still perform optimizations because we can treat RAGs as having an underlying structure. Specifically, while RAG workflows may contain conditional recursion, the core logic remains a Directed Acyclic Graph (DAG). By gleaning this graph structure, a serving system can perform look-ahead resource allocation and optimization without requiring the user to provide a static, exhaustive specification.
Thus, \system can support both the convenience of an imperative programming model and the optimizations present in declarative systems.

Next, we describe how \system{}'s deployment layer uses this structure to map the workflow onto the infrastructure.

\noindent\textbf{Capturing the core RAG pipeline.}
Since our programming model (\S\ref{sec:rag_spec}) allows workflows to be expressed in standard Python rather than a restrictive domain-specific language (DSL), capturing the execution structure is inherently difficult due to Python's dynamic nature~\cite{he2021potential}. However, we observe that for RAG orchestration, we do not require the full-program compilation or low-level C-API interfacing used by high-performance compilers like TorchInductor~\cite{ansel2024pytorch} or Numba~\cite{lam2015numba}. Instead, we only need to extract the logical computational graph—specifically, the dependencies between major RAG components. To achieve this, \system performs a static analysis of the Python Abstract Syntax Tree (AST)\footnote{The AST provides a hierarchical tree representation that maps the structural constructs of the source code, where each node corresponds to a specific programming construct—such as function definitions, loops, or decorators—allowing the system to analyze code logic without executing it.}
We traverse the AST to identify RAG-specific components marked by our decorators (\texttt{harmonia.make} in Figure~\ref{fig:code_listing}), mapping their call sites to derive the underlying graph topology. This coarse-grained approach is intentionally lightweight; it bypasses the complexities of object layout preservation and calling conventions required by traditional compilers, providing just enough structural visibility to enable resource planning and scheduling optimizations discussed later.

\noindent\textbf{Resource Allocation:}
As discussed in \S\ref{sec:background}, RAG components scale differently as configurations change, causing bottlenecks to shift across pipeline variants and over time. Without a systematic approach, developers are left to manually tune resource allocations--a labor‑intensive, error‑prone process that often results in suboptimal utilization.
Building on the observation that the RAG backbone forms a directed graph, we model resource allocation as a variant of the generalized network flow problem (detailed in Figure~\ref{fig:allocation}). Unlike traditional formulations that assume fixed capacities and linear flow behavior, our solution embeds component‑specific scaling--such as batching effects and heterogeneous resource costs--directly into the flow constraints. This is crucial for RAG pipelines, where performance is driven by configuration rather than raw capacity, enabling allocations that accurately reflect real component scaling behavior.

\noindent\underline{\emph{Optimization:}}
Let $G = (V, E)$ be a directed acyclic graph representing a RAG pipeline, where $V$ is the set of component nodes (\eg retrievers, generators) and $E \subseteq V \times V$ is the set of directed edges connecting them. Let $s,t \in V$ denote the source (entry) and sink (exit) nodes, respectively. Let $K$ denote a set of distinct resource types (\eg CPU, GPU).
For each node $i \in V$ and resource type $k \in K$, we define the following quantities. The throughput coefficient $a_{i,k} \geq 0$ represents the throughput capacity that node $i$ achieves per unit of resource $k$. The amplification factor $\gamma_i \geq 0$ captures how much node $i$ amplified the number of requests. The routing probability $p_{i,j} \in [0, 1]$ represents the fraction of node $i$'s output routed to node $j$, subject to $\sum_{j:(i,j)\in E} p_{i,j} = 1$. The total capacity of each resource type is bounded by a budget $C_k \geq 0$.
Given these parameters, the goal is to determine the resource allocation $r_{i,k} \geq 0$, indicating how many units of resource type $k$ to assign to node $i$, such that end-to-end throughput is maximized under the budget constraints.

This extends the standard network flow problem by treating node capacities as endogenous decision variables; rather than optimizing flow across fixed edges, the Gurobi~\cite{gurobi} solver determines the optimal resource assignment $r_{i,k}$ to maximize aggregate throughput at the sink. 
We further constrain the model using profile-driven routing, fixing the proportions $p_{i,j}$ based on empirical data (more below) to ensure the flow respects the data-dependent branches of the RAG application. By incorporating a multi-dimensional resource pool $K$, the optimizer accounts for heterogeneous hardware constraints while simultaneously modeling request amplification ($\gamma_i > 1$), flow abridgement ($\gamma_i < 1$), and the stochastic overhead of recursive loops within a unified framework.

Estimating $\alpha_{i,k}$, $\gamma_{i}$ and $p_{i,j}$ needs profiling, discussed next.

\begin{figure}[t]
\begin{small}
\begin{align*}
\max \; & \sum_{u:(u,t) \in E} f_{ut} && \text{(maximize flow)} \\
\text{s.t. } 
& \sum_{i \in V} r_{i,k} \le C_k \forall k \in K 
  && \text{(resource budget)} \\
&  \sum_{u:(u,i)\in E} f_{ui} \le \sum_{k \in K} \alpha_{i,k} r_{i,k} \forall i \in V 
  && \text{(node capacity constraints)} \\
& f_{i,j} = p_{i,j} \gamma_i \sum_{u:(u,i)\in E} f_{ui} \forall (i,j) \in E
  && \text{(branching and conditionals)} \\
& f_{i,j} \ge 0 \; r_{i,k} \ge 0
  && \text{(nonnegative value)} \\
\end{align*}
\end{small}
\vspace{-20pt}
\caption{\small{Formulation of resource allocation and batch size optimization across a RAG workflow.}}
\vspace{-20pt}
\label{fig:allocation}
\end{figure}

\noindent\underline{\emph{Profiling:}}
\system at startup utilizes a profiling phase to estimate the throughput coefficients $\alpha_{i,k}$, amplification factors $\gamma_i$, and routing proportions $p_{i,j}$. This is achieved by orchestrating the RAG components on representative hardware and executing the pipeline over a subset (\eg random $n=100$ samples) of the ShareGPT~\cite{shareGPT} dataset.

While these offline estimates provide an initial configuration, they are likely to deviate from optimal settings if the actual workload is non-representative or exhibits significant distribution shifts. We discuss how we deal with this sensitivity in \system's runtime layer ($\S$\ref{sec:runtiume_control}). 


\noindent\textbf{Deploying for Distributed Inference.}
To deploy user workflows across distributed setups, at runtime we automatically inject a distribution layer that orchestrates data movement across the infrastructure. 
For multi-node deployments, \system utilizes a high-performance gRPC interface to bridge components; for intra-node execution, it defaults to a low-latency shared-memory substrate to minimize serialization overhead. This approach separates the functional workflow specification from the deployment environment ensuring (i) that developers programs against a simple, logical "single node view" while transparently handling distribution and (ii) the runtime has total visibility and control into inter-component communication—a prerequisite for the fine-grained execution management (\S\ref{sec:runtiume_control}).

This initial optimization pass provides a baseline deployment, but a static allocation is fundamentally insufficient to handle the non-determinism of RAG workloads. Runtime stochasticity—driven by variable component latencies and abrupt shifts in request distributions—renders any single-point optimization a coarse approximation. To mitigate this dynamicity, \system introduces a reactive runtime layer that implements a closed-loop control mechanism. 

\subsection{Run-time Control for RAG inference}
\label{sec:slo_violation}
\label{sec:runtiume_control}
%

To achieve good runtime performance for RAG pipelines, we advocate an SDN-style separation between workflow logic and execution: a centralized control plane coordinates the distributed substrate. 
As illustrated in Figure~\ref{fig:patchwork_overview}, \system's runtime layer is a centralized control plane and request coordinator. The workflow specification and deployment layers together expose a unified control and telemetry surface: the specification layer provides a structured representation of the workflow and its dependencies, while the deployment layer enforces placement and resource decisions. This integration forms a closed loop in which the controller continuously collects execution telemetry (\eg stage latencies and path-dependent control-flow frequencies) and translates these signals into updated routing, scheduling priorities, and resource configurations applied through the deployment substrate. 

The workflow (Figure~\ref{fig:code_listing}) executes under the controller, which drives each request by invoking the appropriate component instances. To keep the controller off the data path, \system{} decouples control from data movement: it schedules stages without ingesting intermediate outputs. For example, after invoking the retriever (Line~34), the controller selects a grader instance (Line~35) and initiates a direct transfer of the retriever output to that grader. 
The controller re-ingests intermediate data only when required by user logic (\eg a conditional check at Line~22). This keeps the controller lightweight but preserves high-bandwidth, low-latency data paths between components.

This centralized decision-making differs from SDN, where the controller installs rules proactively rather than being invoked on every remote call. However, in our experiments(\S\ref{sec:runtime}), we find that decision-making is extremely fast, adding only $\approx$2 ms of overhead (negligible for RAG inference).

With global visibility into request progress and resource pressure, the controller can dynamically rebalance load and reconfigure component instances in response to shifting query mixes and transient hotspots. We describe the runtime policies that drive these decisions next.

\subsubsection{Runtime Mechanisms in \system}
A centralized control plane enables \system{} to implement runtime control mechanisms that existing systems lack.

\noindent
{\bf Resource Reallocation.}
The runtime controller maintains a global view of the system's state, aggregating real-time telemetry on per-component load and the empirical traversal frequency of branches within the pipeline backbone. By monitoring these execution traces, \system continuously re-estimates the critical parameters of the optimization model: component throughput ($\alpha_{i,k}$), request amplification ($\gamma_{i}$), and branching probabilities ($p_{i,j}$). This allows the controller to detect performance drift or workload shifts that invalidate the initial offline profiling. By periodically re-solving the max-flow formulation (\S~\ref{sec:deployment}) with these updated estimates, \system dynamically reconfigure resource allocations in response to transient system pressure. 
In our experiments, we solve the optimization problem in a separate thread every 10 seconds and reallocate resources only when two consecutive solutions agree on the allocation. Owing to the linear formulation, optimization overhead is low -- 3.8 to 31.3ms -- depending on the number of nodes (\S\ref{sec:deploy-eval}).

\noindent
{\bf Load and State Aware Routing.}
Current distributed runtimes~\cite{moritz2018ray} naively dispatch stateless requests to the idle worker. This overlooks the reserved capacity needed for upcoming stateful iterations; an instance that looks "idle" may in fact be a poor candidate if a burst of re-entrant stateful requests is likely to return to it.
With its centralized scheduling approach, \system anticipates future load by tracking each instance's outstanding stateful tasks and their expected re‑entry patterns. The scheduler uses hints derived from annotations and historical execution traces to estimate when stateful iterations will return and how much capacity they will need. Routing decisions account for both current availability and predicted near‑future demand, avoiding overload on instances that only appear idle in the moment.

\noindent
{\bf Communication Granularity Management.} As shown in \S\ref{sec:background} (Figure~\ref{fig:streaming_impact}), naive streaming can reduce throughput under load; \system{} therefore makes 
{\em chunking and streaming granularity load-dependent and runtime-controlled}. Leveraging its global visibility, the runtime controller dynamically modulates the streaming chunk size based on real-time load and pre-profiled component characteristics. To modulate the chunk size, we use offline profiled information on the impact of different chunk sizes at different loads and on request throughput. During online inference, the controller uses this profiled information to change the chunk size.

\subsubsection{SLO Management}

These mechanisms help \system{} improve throughput (\S~4) and achieve high SLO compliance, which current RAG frameworks can't provide. 

Here, we leverage a key architectural insight: {\em while end-to-end latency in iterative or conditional RAG pipelines is often stochastic, individual component latencies exhibit high inter-component correlation.} For example, in a Corrective RAG pipeline, the grader's (an LLM that evaluates the quality of retrieved documents) execution time depends on the volume of retrieved documents, while the retriever time depends on the number of documents to be retrieved (\S\ref{sec:challenges}).

Based on the captured cross-component dependencies, the \system controller dynamically refines its estimate of the remaining execution time for every in-flight request. To achieve this, the runtime maintains online linear regression models that map upstream execution features—such as the number of retrieved documents or token counts—to downstream component latencies. These lightweight models allow \system to estimate the "slack" (the window between the predicted completion time and the SLO deadline) with minimal computational overhead.

Using slack predictions, we implement a deadline-aware scheduler which runs on the runtime controller that dynamically reorders the request execution order at each component, while also routing requests in a load-aware manner. Requests with the least remaining slack are assigned elevated priority. This priority is not just a local scheduling hint; it is also propagated to the managed communication interface, which enforces priority-aware queuing at the network and component layers. This proactive end-to-end management ensures high SLO attainment even under heavy load or significant input-dependent variability, as we show in \S~4.

\subsection{Discussion}

Next, we highlight salient features of \system. 

\noindent\textbf{Applicability on Agentic Serving}
While \system targets structurally heterogeneous RAG pipelines, it does not directly extend to general agentic workflows. Agentic systems exhibit ad hoc, LLM-driven control flow~\cite{team2025kimi}, with task sequences and tool invocations generated dynamically at runtime, violating the static execution backbone assumed by our SLO minimization and max-flow–based resource allocation models. Nonetheless, we believe our core architectural insight—the need for a centralized runtime control plane—remains foundational for agentic serving. We leave support for fully dynamic agent workflows to future work.

\noindent\textbf{Run-Time Scalability.}
The \system's run-time control is central to its ability to handle high load and manage SLO violations. At peak throughput, runtime controller can become a potential bottleneck. To address this, we designed it to scale by leveraging vectorized operations and multi-threading. In our evaluation studies (discussed in Section~\ref{sec:runtime}), we show that even at a high request rate of 1024 requests per second, \system's scheduling latency remains below $2.3$ ms. This level of throughput demonstrates that \system is comparable to prior work~\cite{kwon2023efficient, yu2022orca, infercept}, underscoring its efficiency. 







\section{Evaluation}
\label{ref:eval}

\begin{table}[t]
\scriptsize
\centering
\caption{\small{\textbf{RAG applications and their components:} We provide a brief overview of RAG applications implemented in \system and the structure of workflow.}}
\vspace{-10pt}
\resizebox{0.65\columnwidth}{!}{
\begin{tabular}{@{}lll@{}}
\toprule
Name           & Conditional & Recursive             \\ \midrule
Vanilla-RAG    & \xmark       & \xmark \\
Corrective-RAG & \cmark       & \xmark                 \\
Self-RAG       & \cmark       & \cmark                 \\
Adaptive-RAG   & \cmark       & \cmark                 \\ \bottomrule
\end{tabular}
\label{tab:rag_info}
}
\vspace{-15pt}
\end{table}

We evaluate \system{} on four representative RAG workflows and compare it against commercial engines, including LangChain~\cite{langgraph} and Haystack~\cite{Pietsch_Haystack_the_end-to-end_2019}. Under high load, \system{} improves throughput by up to $2.04\times$ and reduces SLO violations by up to 78.4\% relative to all baselines.

\noindent\textbf{Cluster Setup.}
We conduct our evaluation on four servers, each equipped with two 16-core Intel Xeon Silver 4314 CPUs and 256 GiB of RAM, alongside eight NVIDIA A100 GPUs with 80 GiB of on-device memory. The servers run Ubuntu 22.04.1 with Linux kernel ver. 5.15, using CUDA ver. 12.3.

\noindent\textbf{Implementation.}
\system's core is implemented in over 6K lines of Python code. Its runtime controller coordinates worker processes responsible for executing various RAG components using gRPC~\cite{grpc} as the networking backbone to efficiently handle requests and responses between worker nodes, while also monitoring their status. \system uses Redis at each node to temporarily hold component outputs.
\noindent\textbf{Dataset.} We evaluate \system on a dataset of 3000 chats sampled from the LMSYS-Chat-1M dataset~\cite{zheng2023lmsyschat1m}. Each sample includes a conversation ID, model name, conversation text, language tag, and OpenAI moderation API tag. We simulate arrival rates as a Poisson arrival process. Similar to prior works~\cite{lin2025telerag, tan2025towards}, we create a retriever out of Wiki-Dense Passage retrieval ~\cite{karpukhin-etal-2020-dense}, which contains approximately 21 million passages of up to 100 words. For all requests, we choose $k$, the number of documents retrieved, randomly between $100$ to $300$ which is similar to prior work~\cite{lewis2020retrieval}.

\noindent\textbf{Representative Workflows.}
 We implement four representative RAG applications to test the effectiveness of \system. Table~\ref{tab:rag_info} provides a summary of these applications. We describe their execution pipelines below. 

\noindent
$\bullet$ Vanilla RAG (V-RAG). The simplest RAG pipeline, consisting of a retriever and generator. The retriever performs nearest-neighbor search and returns the top $k$ documents as a single concatenated context. There are no conditionals or recursive steps in this workflow.

\noindent
$\bullet$  Corrective RAG (C-RAG)~\cite{yan2024corrective} 
augments standard RAG with by grading over retrieved documents. The workflow performs initial retrieval, followed by conditional refinement—optionally invoking web search—before final generation, with purely conditional control flow and no recursion.

\noindent
$\bullet$ Self-RAG (S-RAG)~\cite{asai2024self}: 
Self‑RAG augments RAG with a query rewriter and a critic model. After an initial generation, the critic scores the output; low scores trigger query rewriting and recursive re-execution of the pipeline, introducing both conditional branching and recursion.


\noindent
$\bullet$ Adaptive RAG (A-RAG)~\cite{jeong2024adaptive}:
Adaptive RAG introduces path-dependent variability by using a query-complexity classifier to choose an execution branch at runtime. Depending on the input, it follows one of three modes: an LLM-only path with no retrieval for simple queries, a single-pass RAG pipeline for standard requests, or a multi-step iterative RAG loop for complex queries. This workflow includes conditional branching and recursive re-execution over a subgraph.



\begin{figure*}[t]
    \begin{center}
    \includegraphics[width=0.5\textwidth]{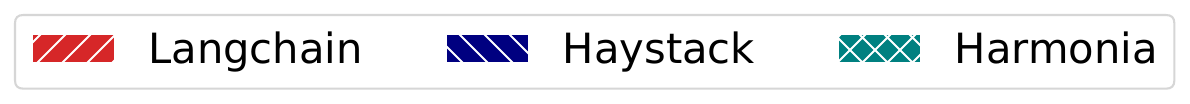}\\
        \begin{subfigure}[b]{0.24\textwidth}
        \includegraphics[width=\linewidth]{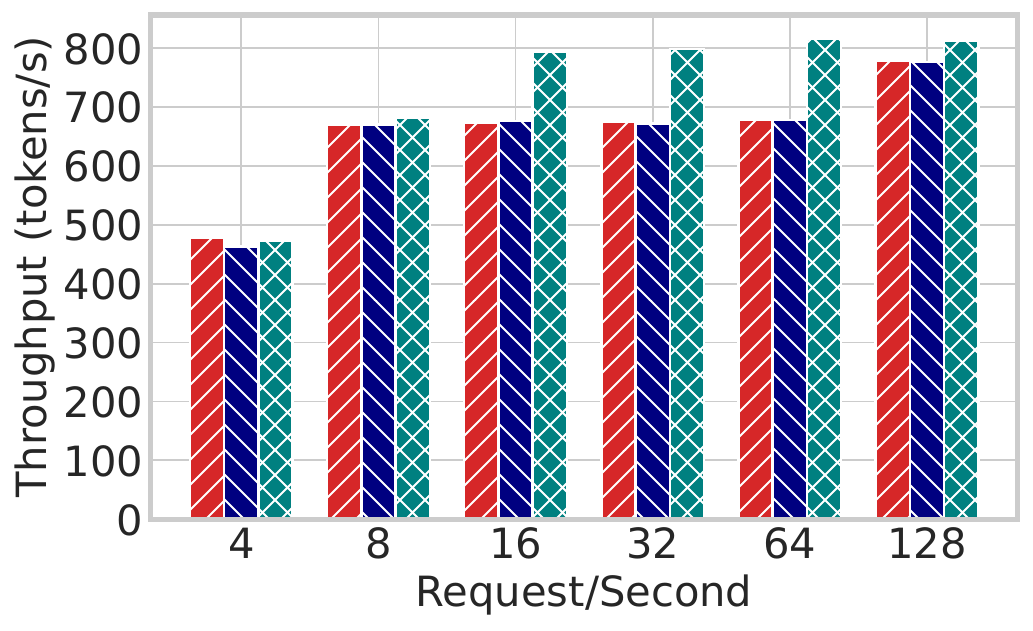}
    \caption{\small{V-RAG 
    }}
    \label{fig:vrag_throughput}
        \end{subfigure}
    \begin{subfigure}[b]{0.24\textwidth}
     \includegraphics[width=\linewidth]{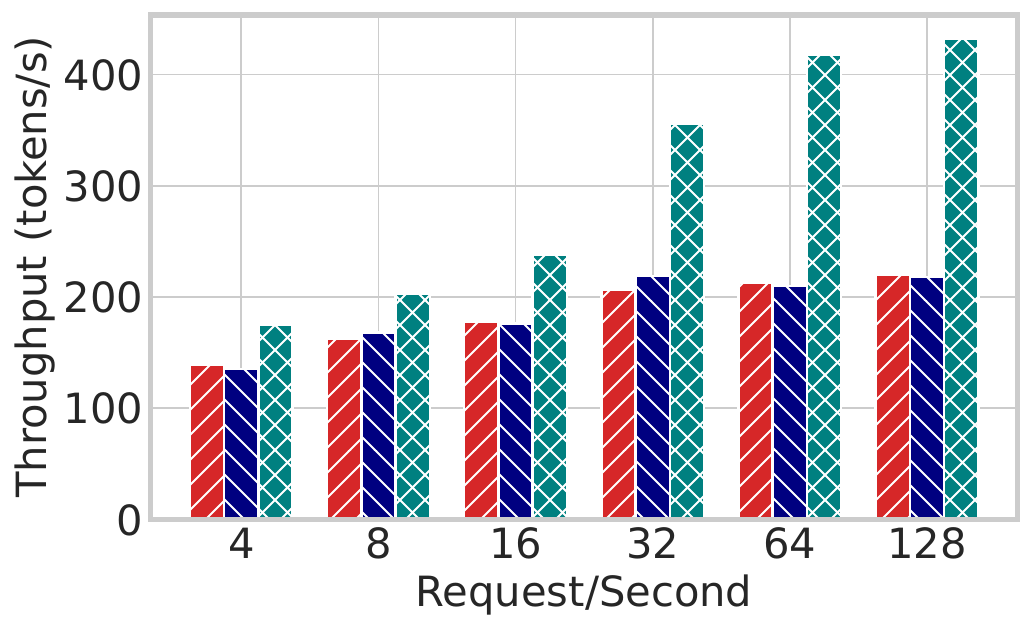}
    \vspace{-10pt}
    \caption{\small{C-RAG}}
    \label{fig:crag_throughput}
    \end{subfigure}
\begin{subfigure}[b]{0.24\textwidth}
 \includegraphics[width=\linewidth]{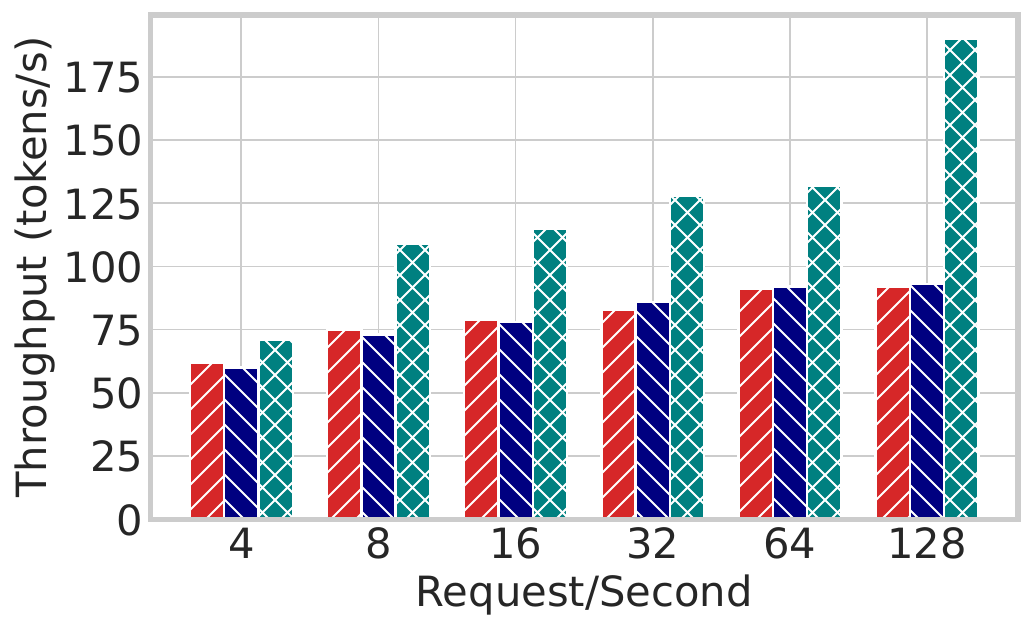}
    \vspace{-10pt}
    \caption{\small{S-RAG}}
    \label{fig:srag_throughput}
    \end{subfigure}
    \begin{subfigure}[b]{0.24\textwidth}
    \includegraphics[width=\linewidth]{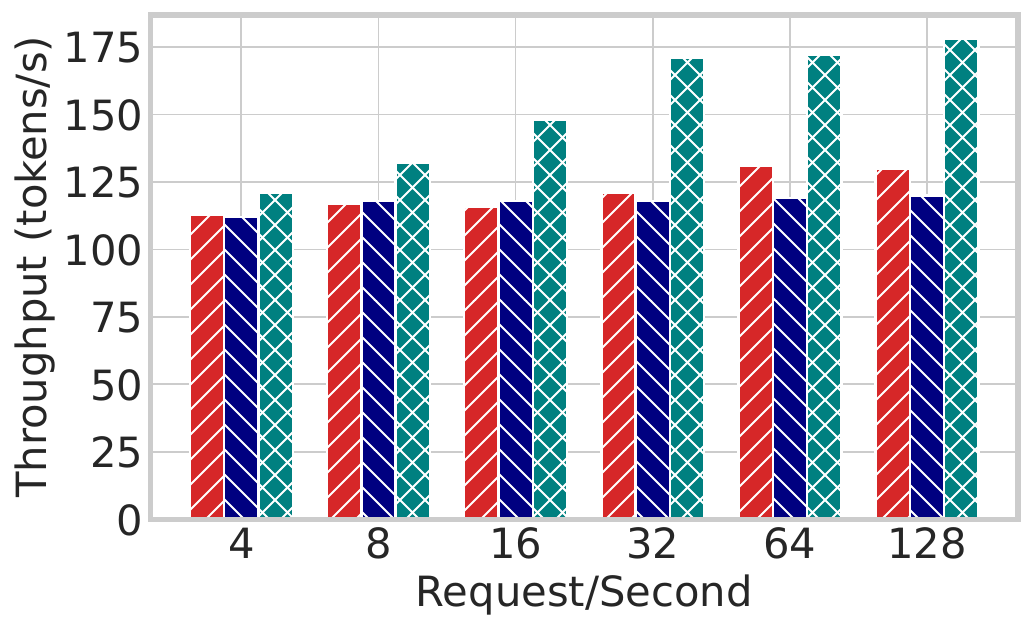}
    \vspace{-10pt}
    \caption{\small{A-RAG}}
    \label{fig:arag_throughput}
    \end{subfigure}
    \end{center}
    \vspace{-1.5em}
    \caption{\small{\textbf{Comparing \system's Throughput:} We observe that \system provides higher throughput of up to $2.04\times$.}}
    \vspace{-10pt}
    \label{fig:system_comparison}
\end{figure*}

\begin{figure}
    \centering
    \includegraphics[width=0.7\linewidth]{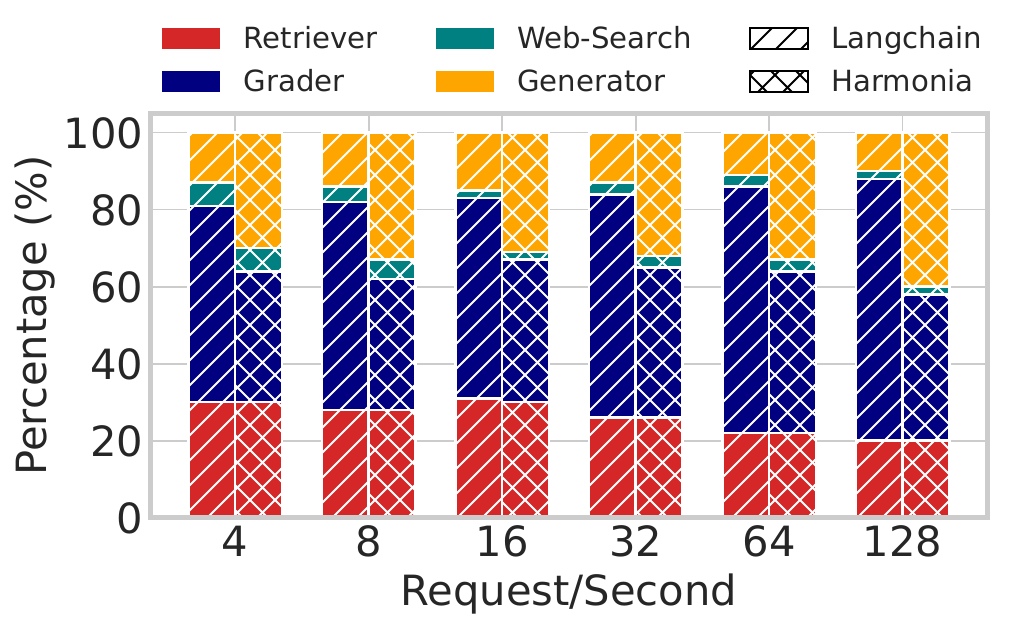}
    \vspace{-10pt}
    \caption{\small{Component Level Breakdown: We observe that CRAG is bottlenecked by the Grader component.
\system{} alleviates it with better
resource allocation.}}
    \label{fig:c-rag-ablation}
    \vspace{-10pt}
\end{figure}

\begin{figure*}[t]
\begin{minipage}[t]{\textwidth}
\begin{center}
    
\vspace{0pt}
    \includegraphics[width=0.5\textwidth]{images/arag_legend.pdf}\\
    
    \begin{subfigure}[b]{0.24\textwidth}
    \centering
        \includegraphics[width=\linewidth]{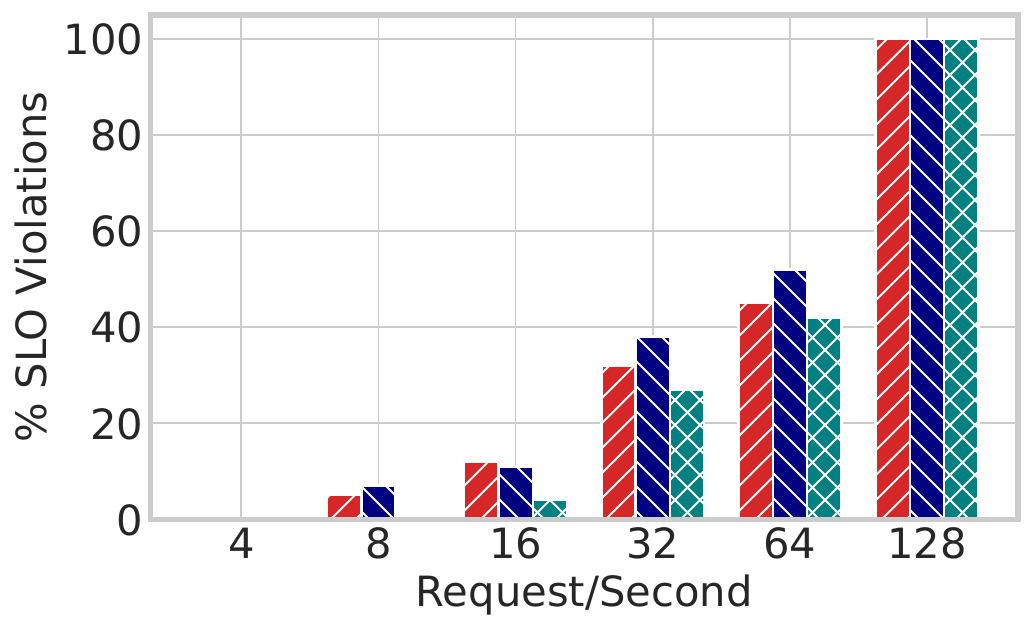}
    \caption{\small{V-RAG}}
    \label{fig:slo-vrag}
        \end{subfigure}
        \begin{subfigure}[b]{0.24\textwidth}
        \centering
       \includegraphics[width=\linewidth]{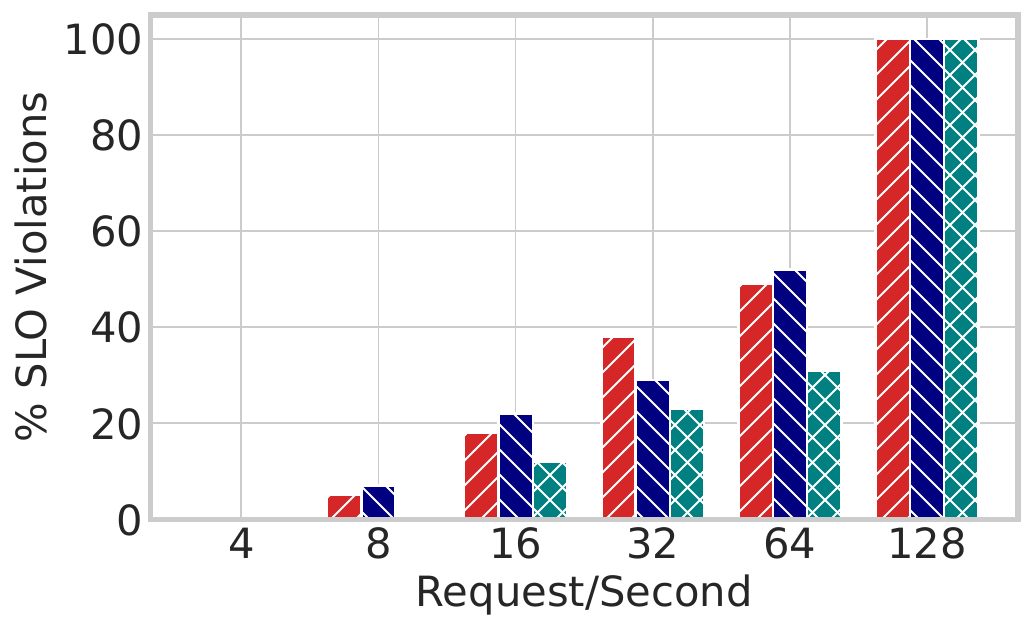}
    \caption{\small{C-RAG
     }}
    \label{fig:slo-crag}
    \end{subfigure}
    \begin{subfigure}[b]{0.24\textwidth}
    \centering
        \includegraphics[width=\linewidth]{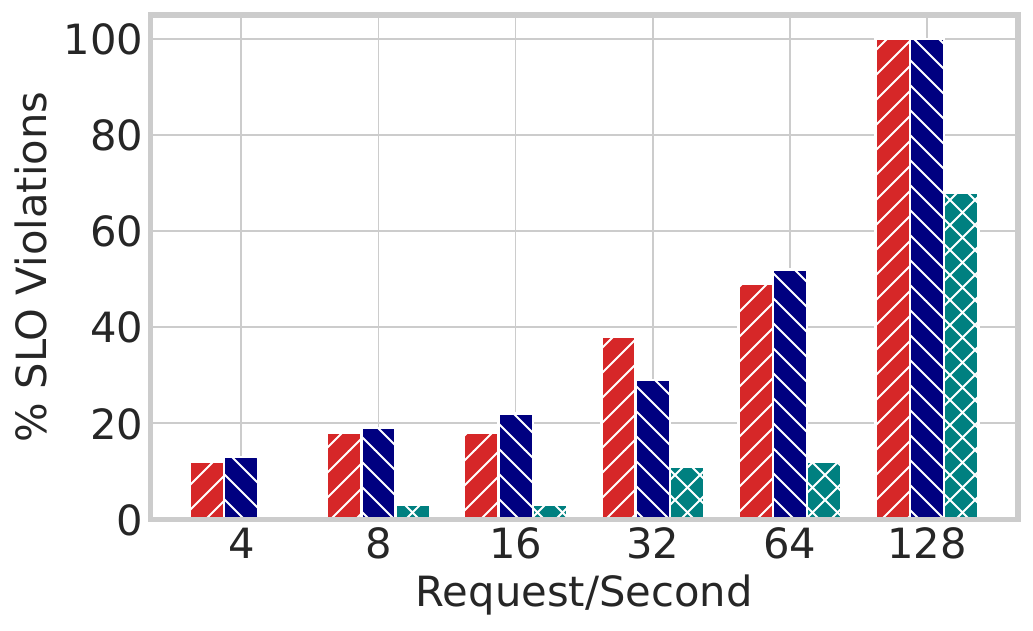}
    \caption{\small{S-RAG}}
    \label{fig:slo-srag}
        \end{subfigure}
        \begin{subfigure}[b]{0.24\textwidth}
        \centering
       \includegraphics[width=\linewidth]{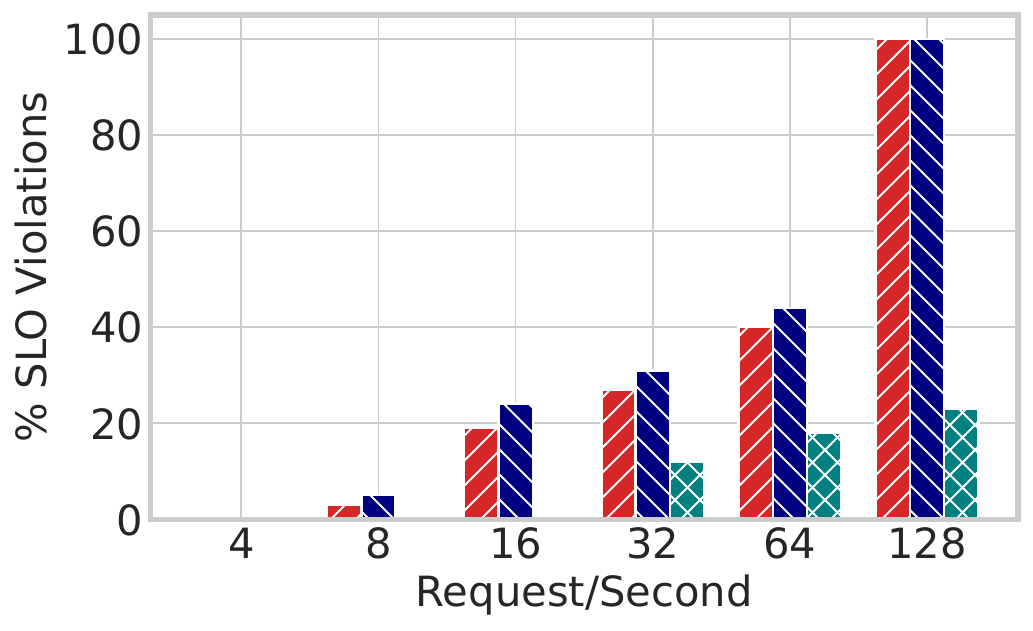}
    \caption{\small{A-RAG}}
    \label{fig:slo_arag}
    \end{subfigure}
    \vspace{-10pt}
    \caption{
    \small{\textbf{SLO Violations: }\system consistently reduces SLO violations when compared to existing serving framework.} 
    }
    \label{fig:slo_violation_minimization}
\end{center}
\end{minipage}
    \vspace{-10pt}
\end{figure*}

\noindent\textbf{Baselines.} 
We evaluate \system against two industry-standard baselines. The first is LangChain~\cite{Chase_LangChain_2022}, a widely adopted development library (over 24k GitHub stars) with its own custom orchestration layer for building and deploying agents. The second is Haystack (over 24k GitHub stars)~\cite{Pietsch_Haystack_the_end-to-end_2019}, which provides a similar modular API but features deep integration with Ray~\cite{moritz2018ray} to manage its distributed runtime.

Note that we have selected a diverse suite of workflows 
that represent the structural complexity and component heterogeneity of modern RAG, including pipelines ranging from simple serial execution to those exhibiting complex conditional branching and recursive control flow. These systems incorporate a wide variety of modules-auxiliary LLMs for query summarization, verification blocks, and external tool calls (\eg web search)—capturing the  heterogeneity of production RAG. To ensure a rigorous comparison, we implement these workflows using the production-grade LangChain and Haystack as competitive baselines.

\subsection{\system's End to End Evaluation}

\label{sec:performnace}

\noindent
{\bf Throughput improvements:}
Figure~\ref{fig:system_comparison} evaluates the throughput of RAG applications implemented in \system against their respective baseline implementations. Across all evaluated scenarios, \system either matches or significantly exceeds the throughput of existing frameworks. For Vanilla RAG—where the workload is a simple DAG with naturally balanced retriever and generator latencies—\system achieves a maximum speedup of approximately 31\%.
However, at near-saturation loads, this advantage narrows to 3\% as all replicas become heavily loaded. For more complex pipelines—specifically C-RAG, S-RAG, and A-RAG—\system improves throughput by up to $1.98\times$, $2.04\times$, and $1.48\times$, respectively, compared to both LangChain and Haystack.

These gains come from heterogeneity-aware provisioning, load-aware routing, and managed streaming, which together reduce stage imbalance. For example, Figure~\ref{fig:c-rag-ablation} shows that \system{} identifies the \emph{Grader} as the bottleneck in C-RAG and allocates additional resources to that stage, improving balance across the pipeline. We quantify the contribution of different \system mechanisms via ablations in \S\ref{sec:runtime}. 

\noindent
{\bf SLO Attainment:}
We now assess the effectiveness of \system's strategy for minimizing SLO violations.
For these experiments, 
we define the SLO threshold as 
2x the average request latency when serving with \system under low load conditions. In Figure~\ref{fig:slo_violation_minimization}, we plot the percentage of SLO violations against the number of incoming requests. 

In general, we see a trend that at moderate loads, \system can reduce SLO violations by 
proactively reordering the request execution.
However, as the system approaches saturation (high load), this advantage diminishes. At these levels, the aggregate request arrival rate exceeds the system's service capacity, causing the slack for all in-flight requests to drop to near zero. In this state, the scheduling problem becomes degenerate; prioritizing one urgent request necessarily causes others to miss their deadlines, which is common in Earliest Deadline First scheduler~\cite{buttazzo1993red}. Consequently, while \system maintains parity with baselines at high load, its prioritization strengths shine during transient bursts, provided there is enough slack to permit effective reordering.

We discuss SLO violation improvement for each RAG next.

\noindent\underline{\emph{V-RAG:}} In Figure~\ref{fig:slo-vrag}, we observe that at moderate loads for V-RAG, \system reduces SLO violations by around 11.8\%. However, at high loads, given that none of the requests have extra slack, and prioritizing deadline first requests ends up delaying other requests, \system shows no improvement. 

\noindent\underline{\emph{C-RAG:}} Compared to LangChain and Haystack, \system shows up to approximately 21\% and 18\% reduction in SLO violations. However, at high load, we face a similar issue to V-RAG where none of the requests have extra slack.

\noindent\underline{\emph{S-RAG:}}
For S‑RAG, \system reduces SLO violations by up to 41.3\% even under high load. We attribute this to execution heterogeneity in S‑RAG: some requests terminate after a single iteration, while others execute multiple recursive loops. This mix creates scheduling slack that \system leverages to prioritize and schedule requests entering their second or third iteration, even at high load.

\noindent\underline{\emph{A-RAG:}} 
For adaptive RAG, requests follow different execution paths, ranging from no retrieval to multi-step retrieval. This execution heterogeneity introduces scheduling slack that \system exploits by prioritizing requests that have already incurred the cost of multi-step retrieval at the generator. As a result, \system reduces SLO violations by up to $78.4\%$ even under high load.

Results indicate that \system{}'s deadline-aware scheduling and slack prediction substantially reduce SLO violations for both static and recursive workflows, enabled by fine-grained, state-dependent prioritization in its centralized controller.
Over the next three subsections, we evaluate each of \system's layers, starting with the specification layer.

\begin{table}[t]
\caption{\small{\textbf{Lines of Code to implement RAG in \system} }}
\centering
\label{tab:loc}
\vspace{-1em}
\resizebox{0.8\linewidth}{!}{
\begin{tabular}{@{}lllll@{}}
\toprule
                                                                      & V-RAG & C-RAG & S-RAG & A-RAG \\ \midrule
\begin{tabular}[c]{@{}l@{}}Abstraction \\ Implementation\end{tabular} & 32    & 78    & 64    & 89    \\ \midrule
Workflow Specification                                                 & 6     & 12    & 14    & 20    \\ \bottomrule
\end{tabular}
}
\vspace{-10pt}
\end{table}

\subsection{Spec. Layer \& Ease of Development} 
\system{} is designed for developer ease. Once developers implement their RAG components, they can deploy the pipeline for distributed inference in $<10$ lines of \system{} code. Table~\ref{tab:loc} summarizes this footprint, reporting the lines of code required to define a RAG application and assemble a complete pipeline using \system{}'s abstractions. Importantly, users specify pipelines as if targeting single-node inference, while \system{} transparently inserts the necessary communication paths and manages distributed deployment. Developers can also write components once and reuse them across pipelines; for example, a vector retriever implemented once can be shared across  RAG workflows.


\begin{figure}[t]
    \centering
    \begin{minipage}[t]{0.46\linewidth}
    \vspace{-5pt}
    \includegraphics[width=\linewidth]{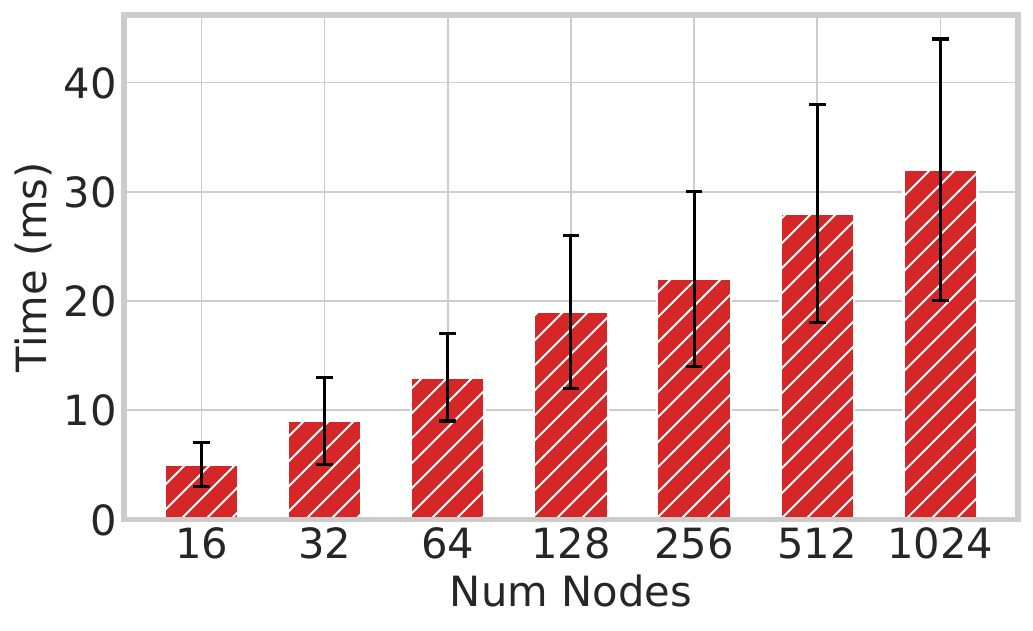}
    \vspace{-15pt}
    \caption{\small{\textbf{Scalability of Optimization Problem:} Optimizer latency with respect to the number of nodes in a RAG application consisting of 16 components.}}
    \label{fig:optimization_scalability}
    \end{minipage}\quad
    \begin{minipage}[t]{0.46\linewidth}
    \vspace{-5pt}
    \includegraphics[width=\linewidth]{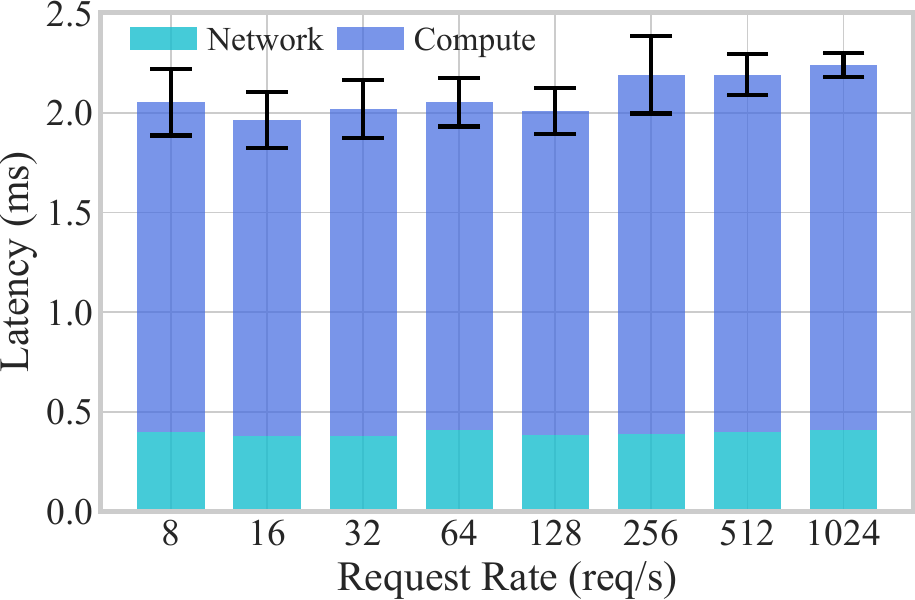}
    \vspace{-15pt}
    \caption{\small{\textbf{Controller processing latency:} 
    Per-request latency remains stable as load increases, indicating that controller does not become a bottleneck.
    }}
    \label{fig:ablation_schd_latency}
    \end{minipage}
    \vspace{-15pt}
\end{figure}

\subsection{Evaluating \system's deployment Layer}
\label{sec:deploy-eval}
We evaluate the scalability of \system's deployment layers by studying the scalability of the optimization problem, understanding the overhead of deployment layer and the resource allocation plans generated by \system.

\noindent\textbf{Scalability of solving the optimization problem.}
\system executes the optimization routine at service launch and re-runs it during auto-scaling events to update batch sizes and resource allocations. The optimization is implemented using Gurobi~\cite{gurobi}. As shown in Figure~\ref{fig:optimization_scalability}, even on a cluster with 1024 nodes, the routine completes in about 
32 milliseconds—demonstrating its efficiency and suitability for dynamic environments. This speed stems from our formulation, which formulates the problem as a linear program (\S\ref{sec:resource_allocation}), ensuring the problem remains linear and tractable. 

\noindent\textbf{Overhead.}
To compare the overhead of \system's deployment layer. We deploy \system on a single node, and compare it to LangChain's single node performance. 
Given that, LangChain utilizes function calls and typically has no overhead associated with cross component communication. 
This enables us to isolate all the overhead \system introduces because of using gRPC. We observe that \system on average only  introduces around 0.8\% overhead over single node performance of LangChain.

\noindent\textbf{Allocation Plans.}
Next, we discuss resource allocation plans generated by \system's deployment layer. 

\noindent\underline{\emph{V-RAG}} For V-RAG, as shown in Figure~\ref{fig:performance_hetero}, the generator and retriever exhibit similar execution times. To maximize throughput, \system launches 8 generators, one on each node, and 8 retrievers, each using 8 CPU cores and 112 GB of RAM.

\noindent\underline{\emph{C-RAG}} For C-RAG, we observe that the grader takes around $1.8\times$ the run-time of the generator. Based on our optimization plan, \system{} selected a configuration with 5 graders and 3 generators, each using 1 GPU. For the retriever, given the graders' low throughput, we launched only 4 retrievers on 4 nodes, each using 8 GPU cores and 112 GB of RAM.

\noindent\underline{\emph{S-RAG}} In S-RAG, we found the generator to be the primary bottleneck as it uses an LLM to generate output based on the retrieved document. The critic, by contrast, has a much higher throughput because it only generates a single output token indicating whether the output is relevant. We observed that \system{} initially started with just 2 critics, 6 generators, and 3 retrievers (on 3 different CPU nodes).

\noindent\underline{\emph{A-RAG}} In A-RAG, we identified the query complexity classifier as the primary bottleneck. 
\system initially deployed 5 query classifiers, 3 generators, and 4 retrievers. During serving, depending on the complexity of incoming queries, we observed the configuration shift to 4 query classifiers and 4 generators. We also observed the number of retrievers fluctuating between 1 and 3, due to high complexity queries.

\begin{table}[t]
\centering
\caption{\small{Throughput (requests/sec) of CPU-heavy retriever (ChromaDB) and GPU-heavy generator (vLLM) when co-located versus running in isolation.}}
\vspace{-10pt}
\resizebox{0.5\columnwidth}{!}{%
\begin{tabular}{lll}
\hline
          & ChromaDB & vLLM \\ \hline
Colocated & 971.9     & 127.6     \\
Isolated  & 972.3     & 128.3     \\ \hline
\end{tabular}%
}
\label{tab:retriever-generator-updated}
\vspace{-20pt}
\end{table}

\noindent\textbf{Co-location} Because \system{} treats CPU, GPU and RAM as heterogeneous components that rely on different resource types can be safely co‑located. \ie a retriever may run on CPU while a generator runs on GPU on the same node without interfering with each other. 
To evaluate the impact of co-location, we deploy ChromaDB and the vLLM inference engine together and measure component-level performance degradation. Table~\ref{tab:retriever-generator-updated} shows that co-location introduces less than $1.1\%$ throughput variance for both components compared to isolated deployment. This minimal overhead is expected, as prior work has demonstrated that CPU resources are rarely the bottleneck in GPU-accelerated inference~\cite{deng2025hgca, vasiliadis2022best}. Slower CPU performance typically does not degrade GPU inference throughput significantly.

\subsection{Evaluating \system's Runtime Layer}
\label{sec:runtime}

\noindent\textbf{Scalability of the \system's runtime controller.}
\system's runtime is responsible for routing, request scheduling, completion tracking, and monitoring. Therefore, its performance can become a bottleneck during scaling.
We evaluate runtime scalability by measuring its latency under varying request rates. As shown in Figure \ref{fig:ablation_schd_latency}, the controller's query-processing latency remains stable across loads, consistently staying around 2 milliseconds. This stability is due to two choices: first, \system's controller operates purely as a control plane, data moves directly between nodes without passing back to the scheduler; second, the critical path control loop is kept minimal, with all resource‑allocation activities handled in a background thread.

\begin{figure}
    \centering
    \includegraphics[width=0.7\linewidth]{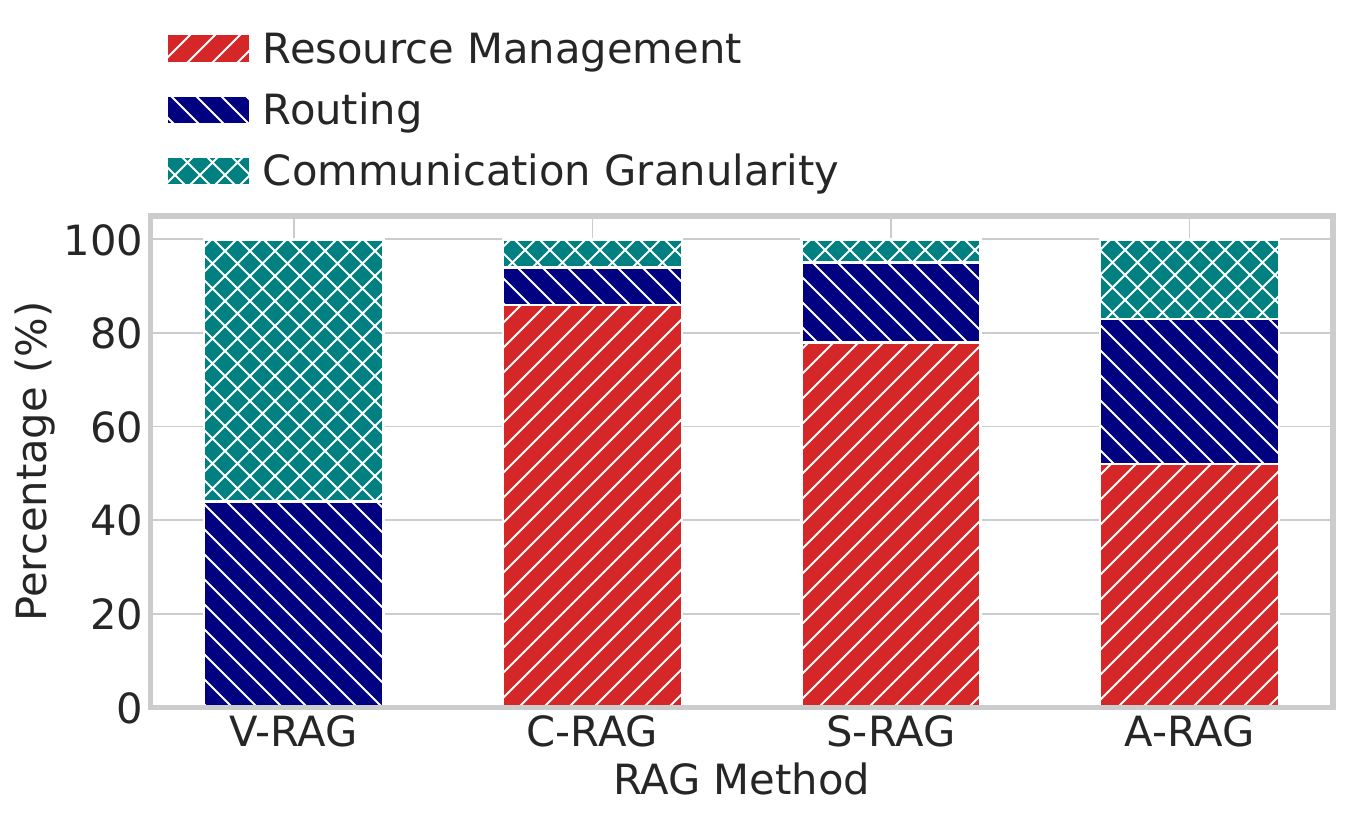}
    \vspace{-10pt}
    \caption{\small{Understanding contributions of each optimization: We plot the contribution of each of our runtime mechanism to achieve speedup for load of 64 reqs/sec.}}
    \label{fig:ablation}
    \vspace{-20pt}
\end{figure}

\noindent\textbf{Contribution of \system's Runtime }
To isolate the individual contributions of \system's runtime mechanisms, we conduct an ablation study in which each mechanism is disabled in turn. We quantify the importance of an optimization as the percentage drop in {\em global throughput} observed when that feature is removed relative to the fully optimized system. Figure~\ref{fig:ablation} shows the proportional contribution of our three primary optimizations across the four evaluated RAG pipelines at a load of 64 requests/sec.

\noindent\underline{\emph{Runtime Resource Management}}
To isolate the contribution of runtime resource management, we fix resource allocations using a one-time static profiling step and disable further tuning. For V-RAG, online resource management provides negligible contribution because the pipeline is static and non‑conditional. In contrast, for C‑RAG, S‑RAG, and A‑RAG, online resource management accounts for 86.8\%, 78.5\%, and 52.1\% of the performance gains, respectively.

\noindent\underline{\emph{Load and State Aware Routing}}
Load- and state-aware routing accounts for $\approx$44\% of the performance gains in V‑RAG. Unlike baselines such as LangChain, which are limited to coarse-grained, top-level replica selection, \system performs fine-grained, component-level routing to the least-loaded instance. For C‑RAG, S‑RAG, and A‑RAG, routing contributes 8.2\%, 17.3\%, and 31.4\% of the overall performance improvements, respectively.

\noindent\underline{\emph{Communication Granularity Management}}
Adjusting communication granularity, defined as the size of streamed chunks, can have a substantial impact on performance. In V-RAG, it accounts for 56.2\% of the overall improvement. This is because fine‑grained streaming enables effective overlap between prefill and decode operations. In other cases, communication granularity management leads to approximately 6\%, 5\%, and 17\% performance improvements, respectively.

No single optimization suffices for RAG serving; \system{} combines complementary mechanisms whose benefits vary across workflows, together ensuring high performance.

\section{Related Work}
\noindent\textbf{Systems for RAGs.}
Several systems improve the scalability of RAG serving by accelerating specific stages or modifying the pipeline. RALMSpec~\cite{zhang2024accelerating} applies speculative execution to reduce RAG latency, and SpeculativeRAG~\cite{wang2024speculative} uses a draft LLM for speculative decoding in generation. Chameleon~\cite{jiang2023chameleon} proposes a heterogeneous architecture to speed up retrieval, while PiperRAG~\cite{jiang2024piperag} introduces a software--hardware co-design for retriever query execution. More recently, TeleRAG~\cite{lin2025telerag} performs speculative retrieval. These approaches either alter the RAG workflow to extract additional efficiency or optimize a single bottleneck component. In contrast, bottlenecks in practice shift across pipelines and configurations, limiting the portability of component-specific optimizations.
RAGO~\cite{jiang2025rago}, offers a taxonomy and abstractions for RAG variants. However, it does not tackle the practical challenges of deployment or resource management required for efficient distributed inference. To the best of our knowledge, \system{} is the first end-to-end RAG serving system that jointly addresses workflow specification, heterogeneity-aware deployment, and closed-loop runtime control.

\noindent\textbf{Comparison with Ray.}
\system{} uses a Python decorator-based interface similar to Ray's, but targets a different execution model tailored to RAG serving. Ray is optimized for general task/actor workloads, whereas RAG pipelines consist of long-running, stateful components requiring sustained provisioning and workflow-aware coordination. Ray's scheduler is event-driven and task-centric: it schedules based on readiness and instantaneous load, with no support for request-level serving semantics such as prioritization or SLO-aware control. In contrast, \system{} treats each query as a first-class request, reasons over the end-to-end workflow graph, and uses this structure to drive placement, routing, and runtime control. We observe \system's design enables to outperform Haystack (which uses Ray) consistently.

\noindent\textbf{Inference systems.}
Prior work on distributed ML inference~\cite{crankshaw2017clipper, gujarati2020serving, romero2021infaas, shen2019nexus} largely targets monolithic models with relatively predictable structure and latency. A separate line of work optimizes LLM serving~\cite{agrawal2023sarathi,kwon2023efficient,yu2022orca,flexgen,NVIDIA_FasterTransformer,hu-etal-2025-stitchllm}, including cache management~\cite{infercept,symphony} and SLO-aware resource adjustment~\cite{romero2021infaas,hu-etal-2024-mosel,shepherd,irina,choi2020lazybatchingslaawarebatchingcloud}. These techniques primarily improve token generation and per-request scheduling, and do not address end-to-end control for heterogeneous RAG pipelines.


\section{Conclusion}
We present \system, an end-to-end system for efficient serving of RAG pipelines. \system introduces modular abstractions that let users express RAG workflows through a high-level, intent-driven Python interface. This framework allows developers to prototype and run applications locally while seamlessly deploying the same code in distributed, production‑scale environments. \system automatically captures the user-defined pipeline, optimizes resource allocation across components, and applies runtime strategies to minimize SLO violations. Together, these optimizations enable \system to deliver up to $2.04\times$ speedup over the best-performing commercial serving systems while reducing SLO violations by as much as $78.4\%$.

\noindent\textbf{Ethics.} {\em This work does not raise any ethical issues.}


\bibliographystyle{plain}
\bibliography{ref}
\clearpage
\appendix
\end{document}